# Security Implications of Distributed Database Management System Models

[1] * Dr.C.Sunil Kumar, [2] J.Seetha, [3] S.R.Vinotha

[1,2,3] *Anna University, Chennai, India*

E-mail: [1] Sunil1977.dce@gmail.com, [2] seetha.venkat80@gmail.com, [3] vinotharamaraj@gmail.com

*Abstract*. Security features must be addressed when escalating a distributed database. The choice between the object oriented and the relational data model, several factors should be considered. The most important of these factors are single and multilevel access controls (MAC), protection and integrity maintenance. While determining which distributed database replica will be more secure for a particular function, the choice should not be made exclusively on the basis of available security features. One should also query the effectiveness and efficiency of the delivery of these characteristics. In this paper, the security strengths and weaknesses of both database models and the thorough problems initiate in the distributed environment are conversed.

* Corresponding address:
Dr.Charupalli Sunil Kumar,
Dhanalakshmi College of Engineering,
Anna University, Chennai, India,
E-mail: Sunil1977.dce@gmail.com, Tel: +91 9498044601

## 1. Introduction

As distributed networks become more accepted, the requirement for improvement in distributed database management systems becomes even more important **[1]**. A distributed system varies from a centralized system in one key respect: The data and often the control of the data are spread out over two or more physically separate locations. Distributed database management systems (DDBMS) are center to many security threats additional to those present in a centralized database management system (CDBMS). In addition, the expansion of sufficient distributed database security has been difficult by the relatively recent opening of the object-oriented database model. This new replica cannot be ignored. It has been created to address the growing difficulty of the data stored in present database systems.

The development of relational database security procedures and standards is a more grown-up field than for the object-oriented model **[9]**. This is mainly due to the fact that object-oriented databases are relatively new **[2]**. The relative immaturity of the object-oriented model is particularly obvious in distributed applications. Incompatible standards are an example: Developers have not embraced a single set of standards for distributed object-oriented databases, while standards for relational databases are well established. One suggestion of this inequality is the inadequacy of controls in multilevel heterogeneous distributed object-oriented systems **[3, 4]** .In this paper; in general will review the security concerns of databases and distributed databases in particular. The security problems found in both models will be examined and the security problems unique to each system will be examined. Finally, the comparison is done relative merits of each model with respect to security.

## *2. DBMS Security Components*
### 2.1 General Database Security Concerns

The distributed database has all of the security apprehensions of a single-location database plus several additional problem areas. The investigation begins with a review of the security elements general to all database systems and those concerns specific to distributed systems. A secure database must satisfy the following requirements.

- It must have substantial integrity (protection from data loss caused by power failures or natural disasters),
- It must have logical integrity (protection of the logical structure of the database),
- It must be obtainable when needed,
- The system must have an review system,
- It must have fundamental integrity (accurate data),
- Access must be controlled to some degree depending on the kindliness of the data,
- A system must be in place to validate the users of the system, and
- Sensitive data must be protected from implication **[8]**.

The following discussion focuses on requirements discussed above, since these security areas are directly affected by the choice of DBMS model. The key objective of these requirements is to ensure that data stored in the DBMS is protected from illegal examination, illegal modification, and from erroneous updates. This can be accomplished by using access controls, concurrency controls, updates using the two-phase commit procedure and inference reduction strategies.

The level of access limit depends on the sensitivity of the data and the degree to which the developer sticks to the principal of least benefit. Typically, a network is maintained in the DBMS that stores the access human rights of individual users. When a user logs on, the interface obtains the specific privileges for the user.

The access permission may be predicated on the satisfaction of one or more of the following criteria:

i. **Availability of data:** Unavailability of data is commonly caused by the locking of a particular data part by another subject, which forces the requesting subject to wait in a queue.
ii. **Adequacy of access:** Only authorized users may analyze and or modify the data. In a single level system, this is relatively easy to implement. If the user is unauthorized, the operating system does not allow system access. On a multilevel system, access control is considerably more difficult to implement, because the DBMS must implement the optional access privileges of the user.
iii. **Assurance of authenticity:** This includes the restriction of access to normal working hours to help ensure that the registered user is authentic. It also includes a usage analysis which is used to determine if the current use is consistent with the needs of the registered user, thereby reducing the probability of an inference attack.

Concurrency controls help to ensure the integrity of the data. These controls normalize the manner in which the data is used when more than one user is using the same data part. These are particularly important in the effective management of a distributed system, because, in many cases, no single DBMS controls data access. If effective concurrency controls are not integrated into the distributed system, several problems can arise. The three possible sources of concurrency problems are identified: (1) Lost update: A successful update was accidentally erased by another user. (2) Unsynchronized transactions that break integrity constraints. (3) Unrepeatable read: Data retrieved is erroneous because it was obtained during an update. Each of these problems can be reduced or eliminated by implementing a suitable locking method or a timestamp method.

    Special problems exist for a DBMS that has multilevel access. In a multilevel access system, users are limited from having complete data access. Policies restricting user access to certain data parts may result from secrecy requirements, or they may result from loyalty to the principal of least privilege (a user only has access to relevant information). Access policies for multilevel systems are typically referred to as either open or closed. In an open system, all the data is considered unclassified unless access to a particular data element is expressly prohibited. A closed system is just the opposite. In this case, access to all data is prohibited unless the user has specific access privileges.

Classification of data parts is not a simple job. This is due, in part, to contradictory goals. The first goal is to provide the database user with access to all non-sensitive data. The second goal is to protect sensitive data from unauthorized inspection. Another problem is data security classification. There is no clear-cut way to classify data. the complexity of the problem are demonstrated: They state that when classifying a data part, there are three proportions:

- The data may be confidential
- The existence of the data may be confidential
- The reason for classifying the data may be confidential

    The first dimension is the easiest to handle. Access to a classified data item is simply starved of. The other two dimensions require more thought and more creative strategies. For example, if an unauthorized user requests a data item whose existence is classified, how does the system respond? A poorly planned response would allow the user to make inferences about the data that would potentially compromise it. Protection from inference is one of the unsolved problems in secure multilevel database design. Several inference protection strategies are listed. These include data suppression, logging every move users make, and perturbation of data. the only practical strategy for the distributed environment that maintains data accuracy is suppression will be discussed later.

### 2.2 Security Problems Unique to DDBMS

- *Centralized or Decentralized Approval*

    In developing a distributed database, one of the first questions to answer is where to grant system access. The two strategies are outlined: (1) Users are granted system access at their home location. (2) Users are granted system access at the remote location. The first case is easier to handle. It is no harder to implement than a centralized access strategy. The success of this strategy depends on reliable communication between the different locations (the remote location must receive all of the necessary authorized information). Since many different locations can grant access, the probability of unauthorized access increases. Once one location has been compromised, the entire system is compromised. If each location maintains access control for all users, the impact of the compromise of a single location is reduced.

    The second strategy, while possibly more secure, has several disadvantages. Probably the most obvious is the additional processing overhead required, particularly if the given process requires the participation of several locations. Furthermore, the maintenance of replicated clearance tables
is computationally expensive and more prone to error. Finally, the replication of passwords, even though they're encrypted, increases the risk of theft. A third possibility is to centralize the granting of access privileges at nodes called policy servers. These servers are arranged in a network. When a policy server receives a request for access, all members of the network determine whether to authorize the access of the user. It is believed that separating the approval system from the application interface reduces the likelihood of compromise.

> *Integrity*

The protection of integrity is much harder in a heterogeneous distributed database than in a homogeneous one. The degree of central control states the level of difficulty with integrity constraints. The homogeneous distributed database has strong central control and has identical DBMS schema. If the nodes in the distributed network are heterogeneous, several problems can occur that will intimidate the integrity of the distributed data. They list three problem areas:

> contradictions between local integrity constraints
> Difficulties in specifying worldwide integrity constraints
> Inconsistencies between local and global limitations

The local integrity constraints are clear to differ in a heterogeneous distributed database. The differences curtail from differences in the individual institutions. These inconsistencies can cause problems, particularly with difficult queries that rely on more than one database. Development of global integrity constraints can abolish conflicts between individual databases. Yet these are not always easy to implement. Global integrity constraints on the other hand are separated from the individual organizations. It may not always be realistic to change the organizational structure in order to make the distributed database consistent. Ultimately, this will lead to inconsistencies between local and global constraints. Conflict resolution depends on the level of central control. If there is strong global control, the global integrity constraints will take priority. If central control is weak, local integrity constraints will.

## 3. Relational Database Protection
### 3.1 Security Concerns
> *Access Controls*

The most ordinary form of access control in a relational database is the view. The view is a logical table, which is created with the SQL VIEW command. This table contains data from the database obtained by additional SQL commands such as JOIN and SELECT. If the database is unclassified, the source for the view is the entire database. If, on the other hand, the database is subject to multilevel classification, then the source for the view is that subset of the database that is at or below the classification level of the user. Users can read or modify data in their view, but the view prohibits users from accessing data at a classification level above their own. In fact, if the view is properly designed, a user at a lower classification level will be unaware that data exists at a higher classification level.

Classification of data and development of legal views become much more complex when the security goal includes the reduction of the threat of inference attacks. Implication is typically made from data at a lower classification level that has been derived from higher level data. The key to this relationship is the derivation rule, which is defined as the operation that creates the derived data. A derivation rule also specifies the access class of the derived data. To reduce the potential for inference, however, the data elements that are inputs to the derivation must be examined to determine whether one or more of these elements are at the level of the derived data. If this is the case, no inference problem exists. If, however, all the elements are at a lower level than the derived data, then one or more of the derivation inputs must be promoted to a higher classification level.

The use of classification constraints to counter implication, beyond the protections provided by the view, requires additional computation. One way that constraint processing can be implemented is discussed. In their model, constraints are processed in three phases. Some constraints are processed during design (these may be updated later), others are processed when the database is queried to authorize access and counter inference, and many are processed during the update phase. Their strategy relies on two implication engines, one for query processing and one for update processing. Essentially, the implication engines are middlemen, which operate between the DBMS and the interface (see Figure 1).

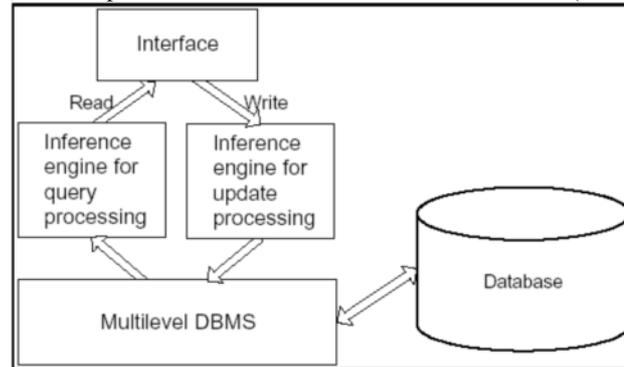

**Figure 1. Constraint processing**

The two implication engines work by evaluating the current task according to a set of rules and determining a course of act. The implication engine for updates dynamically revises the security constraints of the database as the security conditions of the organization change and as the security characteristics of the data stored in the database change. The implication engine for query processing evaluates each entity requested in the query, all the data released in a specific period that is at the security level of the current query, and relevant data available externally at the same security level. This is called the knowledge base. The processor evaluates the potential inferences from the union of the knowledge base and the query's potential response. If the user's security level dominates the security levels of all of the potential inferences, the response is allowed. The integrity constraints in the relational model can be divided into two categories: (1) implied constraints and (2) open constraints. Implied constraints which include domain, relational, and referential constraints enforce the rules of the relational model. Open constraints impose the rules of the organization served by the DBMS. As such, Open constraints are one of the two key elements of security protection in the relational replica.

➢ *Global Views*

As in the centralized relational database, access control in the distributed environment is gifted with the view. Instead of developing the view from local relations, it is developed from the global relations of the distributed database. Accordingly, it is referred to as a global view. The view mechanism is even more important in the distributed environment because the problem is typically more complex and while centralized databases may not be maintained as multilevel access systems, a distributed database is more likely to require the control of information.

Although global views are successful at data control and to a lesser extent at implication protection, their use can be computationally costly. One of the key problems with a relational distributed database is the computation required to execute a complex query. Since each view is unique, a different query is necessary for each view. This additional transparency is partially balance by query optimizers. Nonetheless, the addition of global views adds computing time to a process that already takes too long.

In an effort to provide additional implication protection beyond the global view, it extends their classification constraint processing model to the distributed environment. As with the centralized model, implication engines are added to the standard distributed database architecture at each location. Their model assumes that the distributed database is homogeneous. In this case, the implication engines at the user's location processes the query and update constraints. Only a small amount of overhead is added. If the distributed database is heterogeneous, however, then the processing overhead would be prohibitively expensive since the implication engines at each location involved in the action would need to process the security constraints for all the local data. Considering the processing demands already in place in a relational database management system (RDBMS), this appears to be impractical.

## 4. Object-Oriented Database Security

### 4.1 Object-Oriented Databases

While it is not the aim of this paper to present a detailed description of the object-oriented model, the reader may be unfamiliar with the fundamentals of an object-oriented database. For this reason, a brief look at the object-oriented model's basic structure is viewed. The basic element of an object-oriented database is the object. An object is defined by a class. In essence, classes are the blueprints for objects. In the object-oriented model, classes are arranged in a hierarchy. The root class is found at the top of the hierarchy. This is the parent class for all other classes in the model. We say that a class that is the descendent from a parent *inherits* the properties of the parent class. As needed, these properties can be modified and extended in the descendent class.

An object is composed of two basic elements: variables and methods. An object holds three basic variables types: (1) Object class: This variable keeps a record of the parent class that defines the object. (2) Object ID (OID): A record of the specific object instance. The OID is also kept in an OID table. The OID table provides a map for finding and accessing data in the object-oriented database. It is seen that it also has special significance in creating a secure database. (3) Data stores: These variables store data in much the same way that attributes store data in a relational tuple.

Methods are the actions that can be performed by the object and the actions that can be performed on the data stored in the object variables. Methods perform two basic functions: They communicate with other objects and they perform reads and updates on the data in the object. Methods communicate with other objects by sending messages. When a message is sent to an object, the receiving object creates a subject. Subjects execute methods; objects do not. If the subject has suitable clearance, the message will cause the subject to execute a method in the receiving object. Often, when the action at the called object ends, the subject will execute a method that sends a message to the calling object indicating that the action has ended.

Methods perform all reading and writing of the data in an object. For this reason, the data is *encapsulated* in the object. This is one of the important differences between object-oriented and relational databases. All control for access, modification, and integrity start at the object level. For example, if no method exists for updating a particular object's variable, then the value of that variable is constant. Any change in this condition must be made at the object level. See **Figure 2** for a schematic view of the object-oriented model.

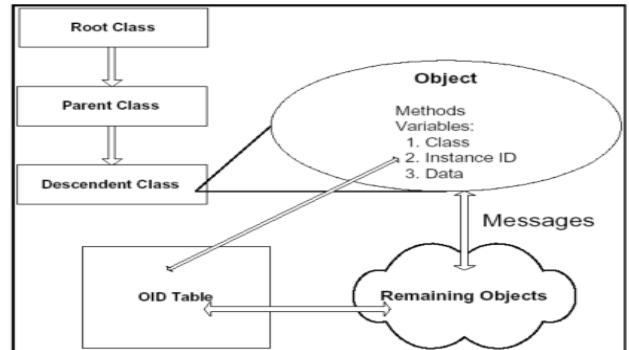
**Figure 2. The object-oriented model.**

**4.2 Security Issues**
➢ *Access Controls*

As with the relational model, access is controlled by classifying elements of the database. The basic element of this classification is the object. Access permission is granted if the user has sufficient security clearance to access the methods of an object. It describes a security model that effectively explains the access control concepts in the object-oriented model. Their model is based on six security properties:

**Property 1** (Hierarchy Property). The level of an object must control that of its class object.
**Property 2** (Subject Level Property). The security level of a subject dominates the level of the invoking subject and it also dominates the level of the home object.
**Property 3** (Object Locality Property). A subject can execute methods or read or write variables only in its home object.
**Property 4** (Property) A subject may write into its home object only if its security is equal to that of the object.
**Property 5** (Return value property) A subject can send a return value to its invoking subject only if it is at the same security level as the invoking subject.
**Property 6** (Object creation property) the security level of a newly-created object dominates the level of the subject that requested the creation.

*4.2.1 Integrity*
As with classification constraints, integrity constraints are also executed at the object level. These constraints are similar to the explicit constraints used in the relational model. The difference is in execution. An object-oriented database maintains integrity before and after an update by executing constraint checking methods on the affected objects. A relational DBMS takes a more global approach **[5]**. One of the benefits of encapsulation is that subjects from remote objects do not have access to a called object's data. This is a real advantage that is not present in the relational DBMS. It is noted that an object oriented system derives a significant benefit to database integrity from encapsulation. This benefit stems from modularity. Since the objects are encapsulated, an object can be changed without affecting the data in another object. So, the process that contaminated one element is less likely to affect another element of the database.

There are many obstacles to the successful implementation of a distributed object-oriented database. The association of the object-oriented DDBMS is more complex than the relational DDBMS. In a relational DDBMS, the role of client and server is maintained. This makes the development of multilevel access controls easier. Since the roles of client and server are not well defined in the object-oriented model, control of system access and multilevel access is more difficult. System access control for the object-oriented DDBMS can be handled at the host location in a procedure similar to that described for the relational DDBMS. Since there is no clear definition of client and server, however, the use of replicated multisite approval would be impractical. Multilevel access control problems arise when developing effective and efficient authorization algorithms for subjects that need to send messages to multiple objects across several geographically separate locations. There are currently no universally accepted means for enforcing subject authorization in a pure object-oriented distributed environment.

This means that, while individual members have developed their own authorization systems, there is no pure object-oriented vendor-independent standard which allows object-oriented database management systems (OODBMS) from different vendors (a heterogeneous distributed system) to communicate in a secure manner. Without subject authorization, the controls described in the previous section cannot be enforced. Since inheritance allows one object to inherit the properties of its parent, the database is easily compromised. So, without effective standards, there is no way to enforce multilevel classification. It is noted that one standard does exist, called OSF DCE (Open Software Foundation's Distributed Computing Environment), that is vendor-independent, but is not strictly an object-oriented database standard. While it does provide subject authorization, it treats the distributed object environment as a client/server environment as is done in the relational model. The major integrity concern in a distributed environment that is not a concern in the centralized database is the distribution of individual objects **[6]**. Recall that a RDBMS allows the fragmentation of tables across sites in the system. It is less desirable to allow the fragmentation of objects because this can break encapsulation. For this reason, fragmentation should be explicitly forbidden with an integrity constraint.

## 6. Conclusion

Database security issues are discussed in general and how the database model affects database system security in particular. It has seen that security protections for OODBMS and RDBMS are quite different. Each replica has important strengths and weaknesses. Currently, the RDBMS is the better choice for a distributed application. This is due to the relative maturity of the relational model and the existence of universally accepted standards. The recent emergences of mixture models that combine the features of the two models discussed raise many new security questions. For example, Informix's Illustra combines a relational database schema with the capability to store and query complex data types **[7]**. They call this system an "object-relational database." They include subject authorization strategies for heterogeneous distributed systems, inference prevention strategies for both centralized and distributed database systems, and distributed object-oriented database security standards.

### Acknowledgements

The author would like to express their sincere gratitude to the Management of **Dhanalakshmi College of Engineering, Chennai** for their constant encouragement and co-operation.

## Biography

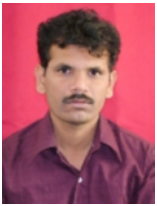

*Dr. C.Sunil Kumar* did his **B.E (CSE)** degree from **University of Madras**, in 1998, **M.Tech (CSE)** degree from **SRM University**, Chennai, India, in 2005 and **Ph.D (C.S.E)** from **JNTUH, Kukatpally**, Hyderabad in 2012 in the area of **Distributed Databases: a Case Study on Healthcare Information Systems.** Currently, he is **Professor** & **HOD** in **Computer Science & Engineering Department, Dhanalakshmi college of Engineering, Chennai, India.** He has guided more than 40 B.Tech and M.Tech projects. He has contributed 2**2 Research Publications** at International/National Journals and Conferences. His research interests are Distributed Databases, Object Oriented Technologies and Information

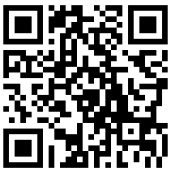

Free download this article
and more information